\documentclass[11pt]{article}
\usepackage{jheppub}
\usepackage{axodraw}
\usepackage{psfrag} 
\usepackage{bbm} 
\allowdisplaybreaks 

\newcommand{\la}[1]{\label{#1}}
\newcommand{\rmi}[1]{{\mbox{\scriptsize #1}}}
\newcommand{\fig}{Fig.~}
\newcommand{\eq}{Eq.~}
\newcommand{\se}{Sec.~}
\newcommand{\eqs}{Eqs.~}
\newcommand{\nr}[1]{(\ref{#1})}
\newcommand{\nn}{\nonumber\\}
\newcommand{\fr}[2]{{\frac{#1}{#2}\,}}
\renewcommand{\vec}[1]{{\textbf{#1}}}
\renewcommand{\(}{\left(}
\renewcommand{\)}{\right)}
\newcommand{\lb}{\left\{}
\newcommand{\rb}{\right\}}
\newcommand{\lk}{\left[}
\newcommand{\rk}{\right]}
\newcommand{\e}{\epsilon}
\newcommand{\order}[1]{\mathcal{O}\!\(#1\)\vphantom{\fr12}}
\newcommand{\sy}[3]{{\textstyle #1\frac{#2}{#3}}} 
\newcommand{\sm}[1]{\,{\scriptstyle #1}\,} 
\newcommand{\sumint}[1]{\hbox{$\sum$}\!\!\!\!\!\!\!\int_{#1}}
\newcommand{\sumintp}[1]{\hbox{$\sum^\prime$}\!\!\!\!\!\!\!\!\!\int_{#1}}

\newcommand{\intr}{\int_0^\infty\!\!\!\!{\rm d}r\,}
\newcommand{\li}[1]{{\rm Li}_{#1}(e^{-2r})}
\newcommand{\gammaE}{\gamma_{\small\rm E}}
\newcommand{\CA}{C_{\mathrm{A}}}
\newcommand{\CF}{C_{\mathrm{F}}}
\newcommand{\Nf}{N_{\mathrm{f}}}
\newcommand{\Nc}{N_{\mathrm{c}}}

\newcommand{\II}{I}
\newcommand{\Ib}[2]{\II_{#1}^{#2}}
\newcommand{\If}[2]{\hat \II_{#1}^{#2}}
\newcommand{\JJ}[2]{J_{#1}^{#2}}
\newcommand{\KK}[2]{K_{#1}^{#2}}
\newcommand{\LL}[2]{L_{#1}^{#2}}
\def\PiEa{\Pi_{\mathrm{E1}}(0)} \def\PiEap{\Pi_{\mathrm{E1}}'(0)} \def\PiEapp{\Pi_{\mathrm{E1}}''(0)}
\def\PiEb{\Pi_{\mathrm{E2}}(0)} \def\PiEbp{\Pi_{\mathrm{E2}}'(0)}
\def\PiEc{\Pi_{\mathrm{E3}}(0)} 
\def\PiTa{\Pi_{\mathrm{T1}}(0)} \def\PiTap{\Pi_{\mathrm{T1}}'(0)} \def\PiTapp{\Pi_{\mathrm{T1}}''(0)}
\def\PiTb{\Pi_{\mathrm{T2}}(0)} \def\PiTbp{\Pi_{\mathrm{T2}}'(0)}
 \def\PiTcp{\Pi_{\mathrm{T3}}'(0)}
\newcommand{\PiE}{\Pi_{\mathrm{E}}}
\newcommand{\PiT}{\Pi_{\mathrm{T}}}
\newcommand{\PiEQCD}{\Pi_{\mathrm{EQCD}}}
\newcommand{\gE}{g_{\mathrm{E}}}
\newcommand{\gM}{g_{\mathrm{M}}}
\newcommand{\mE}{m_{\mathrm{E}}}
\newcommand{\mel}{m_{\mathrm{el}}}
\newcommand{\lE}{\lambda_{\mathrm{E}}}
\newcommand{\Pic}[4]{\;\parbox[c]{#2 pt}{\begin{picture}(#2,#3)(0,0)
\SetWidth{1.0}\SetScale{#4} #1 \end{picture}}\;}
\newcommand{\pic}[1]{\Pic{#1}{30}{30}{1.0}}
\newcommand{\picb}[1]{\Pic{#1}{45}{30}{1.0}}
\renewcommand{\pic}[1]{\Pic{#1}{21}{21}{0.7}}
\renewcommand{\picb}[1]{\Pic{#1}{31.5}{21}{0.7}}

\def\Agl(#1,#2)(#3,#4,#5){\PhotonArc(#1,#2)(#3,#4,#5){1}
{6.283 #3 mul 360 div #4 #5 sub #4 #5 sub mul sqrt mul Ldensity mul}}
\def\Lgl(#1,#2)(#3,#4){\Photon(#1,#2)(#3,#4){1}
{#1 #3 sub #1 #3 sub mul #2 #4 sub #2 #4 sub mul add sqrt Ldensity mul}}
\def\Legl(#1,#2)(#3,#4){\Photon(#1,#2)(#3,#4){1.25}
{#1 #3 sub #1 #3 sub mul #2 #4 sub #2 #4 sub mul add sqrt Ldensity mul 0.78 mul}}
\def\Agh(#1,#2)(#3,#4,#5){\DashArrowArc(#1,#2)(#3,#4,#5){1}}
\def\Aagh(#1,#2)(#3,#4,#5){\DashArrowArcn(#1,#2)(#3,#5,#4){1}}
\def\Lgh(#1,#2)(#3,#4){\DashArrowLine(#1,#2)(#3,#4){1}}
\def\Lagh(#1,#2)(#3,#4){\DashArrowLine(#3,#4)(#1,#2){1}}
\def\Ahh(#1,#2)(#3,#4,#5){\DashCArc(#1,#2)(#3,#4,#5){1}}
\def\Lhh(#1,#2)(#3,#4){\DashLine(#1,#2)(#3,#4){1}}
\def\Aqu(#1,#2)(#3,#4,#5){\ArrowArc(#1,#2)(#3,#4,#5)}
\def\Aaqu(#1,#2)(#3,#4,#5){\ArrowArcn(#1,#2)(#3,#5,#4)}
\def\Lqu(#1,#2)(#3,#4){\ArrowLine(#1,#2)(#3,#4)}
\def\Laqu(#1,#2)(#3,#4){\ArrowLine(#3,#4)(#1,#2)}
\def\Aqq(#1,#2)(#3,#4,#5){\CArc(#1,#2)(#3,#4,#5)}
\def\Lqq(#1,#2)(#3,#4){\Line(#1,#2)(#3,#4)}
\def\Asc(#1,#2)(#3,#4,#5){\CArc(#1,#2)(#3,#4,#5)}
\def\Lsc(#1,#2)(#3,#4){\Line(#1,#2)(#3,#4)}
\def\TopoVR(#1){\pic{#1(15,15)(15,-90,270)}}
\def\ToptVS(#1,#2,#3){\pic{#1(15,15)(15,0,180) #2(15,15)(15,180,360)%
 #3(30,15)(0,15)}}
\def\ToprVM(#1,#2,#3,#4,#5,#6){\pic{#3(15,15)(15,-30,90) #1(15,15)(15,90,210)%
 #2(15,15)(15,210,330) #5(2,7.5)(15,15) #6(15,15)(15,30) #4(28,7.5)(15,15)}}
\def\ToprVV(#1,#2,#3,#4,#5){\!\!\picb{#2(26.25,15)(15,256,76)%
 #3(30,30)(15,30) #1(18.75,15)(15,104,284) #4(15,30)(22.5,0)%
 #5(30,30)(22.5,0)}\!\!}
\def\ToprVB(#1,#2,#3,#4){\picb{#1(30,15)(15,-120,120) #2(30,15)(15,120,240)%
 #3(15,15)(15,60,300) #4(15,15)(15,-60,60)}}
\def\SPropCircProp(#1,#2){\picb{#1(0,15)(15,15) #1(30,15)(45,15)
 \GCirc(22.5,15){7.5}{0.75} \Text(15.75,10.5)[c]{$\scriptstyle #2$}}}
\def\TopoS(#1){\SPropCircProp(#1,1)}
\def\TopoSB(#1,#2,#3){\picb{#1(0,15)(7.5,15) #2(22.5,15)(15,0,180)%
 #3(22.5,15)(15,180,360) #1(37.5,15)(45,15)}}
\def\TopoST(#1,#2){\picb{#1(0,0)(22.5,0) #1(22.5,0)(45,0)%
 #2(22.5,15)(15,-90,270)}} 
\def\ToptS(#1){\SPropCircProp(#1,2)}
\def\ToptSM(#1,#2,#3,#4,#5,#6){\picb{#1(0,15)(7.5,15) #1(37.5,15)(45,15)%
 #2(22.5,15)(15,0,90) #3(22.5,15)(15,90,180) #4(22.5,15)(15,180,270)%
 #5(22.5,15)(15,270,360) #6(22.5,30)(22.5,0)}}
\def\ToptSAl(#1,#2,#3,#4,#5){\picb{#1(0,15)(7.5,15) #1(37.5,15)(45,15)%
 #2(22.5,15)(15,0,90) #3(22.5,15)(15,90,180) #4(22.5,15)(15,180,360)%
 #5(7.5,30)(15,270,360)}}
\def\ToptSAr(#1,#2,#3,#4,#5){\picb{#1(0,15)(7.5,15) #1(37.5,15)(45,15)%
 #2(22.5,15)(15,0,90) #3(22.5,15)(15,90,180) #4(22.5,15)(15,180,360)%
 #5(37.5,30)(15,180,270)}}
\def\ToptSE(#1,#2,#3,#4,#5){\picb{#1(0,15)(7.5,15) #1(37.5,15)(45,15)%
 #3(15,15)(7.5,0,180) #4(15,15)(7.5,180,360)%
 #2(30,15)(7.5,0,180) #5(30,15)(7.5,180,360)}} 
\def\ToptSS(#1,#2,#3,#4){\picb{#1(0,15)(7.5,15) #1(37.5,15)(45,15)%
 #4(7.5,15)(37.5,15) #2(22.5,15)(15,0,180) #3(22.5,15)(15,180,360)}}
\def\ToptSBB(#1,#2,#3,#4,#5){\picb{#1(0,15)(7.5,15)  #1(37.5,15)(45,15)%
 #2(22.5,15)(15,0,70) #2(22.5,15)(15,110,180) #3(22.5,15)(15,180,360)%
 #4(22.5,30)(5,-10,190) #5(22.5,30)(5,190,350)}}
\def\ToptSBT(#1,#2,#3,#4){\picb{#1(0,15)(7.5,15)  #1(37.5,15)(45,15)%
 #2(22.5,15)(15,0,90) #2(22.5,15)(15,90,180) #3(22.5,15)(15,180,360)%
 #4(22.5,35)(5,-90,270)}}
\def\ToptSTB(#1,#2,#3,#4){\picb{#1(0,0)(22.5,0) #1(22.5,0)(45,0)%
 #2(22.5,15)(15,-90,70) #2(22.5,15)(15,110,270)%
 #3(22.5,30)(5,-10,190) #4(22.5,30)(5,190,350)}}
\def\ToptSTT(#1,#2,#3){\picb{#1(0,0)(22.5,0) #1(22.5,0)(45,0)%
 #2(22.5,15)(15,-90,90) #2(22.5,15)(15,90,270)%
 #3(22.5,35)(5,-90,270)}}
\def\ToprS(#1){\SPropCircProp(#1,3)}
\def\ToprSMi(#1,#2,#3,#4,#5,#6,#7,#8){\picb{#1(0,15)(7.5,15) #1(37.5,15)(45,15)%
 #2(22.5,15)(15,0,90) #3(22.5,15)(15,90,180) #4(22.5,15)(15,180,270)%
 #5(22.5,15)(15,270,360) #6(22.5,30)(22.5,16) #7(22.5,16)(11.5,5) #8(22.5,16)(33.5,5)}}
\def\ToprSMiV(#1,#2,#3,#4,#5,#6,#7){\picb{#1(0,15)(7.5,15) #1(37.5,15)(45,15)%
 #2(22.5,15)(15,0,90) #3(22.5,15)(15,90,180) #4(22.5,15)(15,180,270)%
 #5(22.5,15)(15,270,360) #6(22.5,0)(11.5,25) #7(22.5,0)(33.5,25)}}
\def\ToprSMiNp(#1,#2,#3,#4,#5,#6,#7,#8){\picb{#1(0,15)(7.5,15) #1(37.5,15)(45,15)%
 #2(22.5,15)(15,0,90) #3(22.5,15)(15,90,180) #4(22.5,15)(15,180,270)%
 #5(22.5,15)(15,270,360) #6(33.5,5)(11.5,25) #7(26,16)(33.5,25) #8(11.5,5)(20,13)}}
\def\ToprSMiB(#1,#2,#3,#4,#5,#6){\picb{#1(0,15)(7.5,15) #1(37.5,15)(45,15)%
 #2(22.5,15)(15,0,90) #3(22.5,15)(15,90,180) #4(22.5,15)(15,180,360)%
 #5(7.5,30)(15,270,360) #6(37.5,0)(15,90,180)}}
\def\ToprSMiBT(#1,#2,#3,#4,#5,#6,#7){\picb{#1(0,15)(7.5,15) #1(37.5,15)(45,15)%
 #2(22.5,15)(15,0,90) #3(22.5,15)(15,90,180) #4(22.5,15)(15,180,360)%
 #5(22.5,30)(22.5,15) #6(22.5,15)(22.5,0) #7(22.5,15)(37,15)}}
\def\ToprSMiBB(#1,#2,#3,#4,#5,#6,#7){\picb{#1(0,15)(7.5,15) #1(37.5,15)(45,15)%
 #2(22.5,15)(15,0,90) #3(22.5,15)(15,90,180) #4(22.5,15)(15,180,360)%
 #5(22.5,15)(5,0,360) #6(22.5,30)(22.5,20) #7(22.5,10)(22.5,0)}}

\title{Three-loop matching coefficients for hot QCD: Reduction and 
gauge independence}

\preprint{BI-TP 2012/25}

\author{J. M\"oller}
\author{and Y. Schr\"oder}

\affiliation{Faculty of Physics, University of Bielefeld, D-33501 
Bielefeld, Germany}

\emailAdd{jmoeller@physik.uni-bielefeld.de}
\emailAdd{yorks@physik.uni-bielefeld.de}

\abstract{We perform an integral reduction for the 3-loop effective gauge coupling and screening mass 
of QCD at high temperatures, defined as matching coefficients appearing in the 
dimensionally reduced effective field theory (EQCD). 
Expressing both parameters in terms of a set master (sum-) integrals,
we show explicit gauge parameter independence.
The lack of suitable methods for solving the comparatively large number of master integrals
forbids the complete evaluation at the moment. 
Taking one generic class of masters as an example, 
we highlight the calculational techniques involved.
The full result would allow to improve on one of the classic probes for the convergence 
of the weak-coupling expansion at high temperatures, namely the comparison of full
and effective theory determinations of the spatial string tension. 
Furthermore, the full result would also allow to determine one new contribution
of order $\order{g^7}$ to the pressure of hot QCD.}

\begin{document}
\maketitle
\flushbottom

%
\section{Introduction}

Thermal QCD at high temperatures ($T$) exhibits three different momentum 
scales. 
It has been known \cite{Ginsparg:1980ef,Appelquist:1981vg} for a long 
time that the ``soft'' static color-electric modes $p \sim gT$, 
where $g$ is the gauge
coupling, are responsible for the slow convergence whereas the 
``ultra-soft'' static color-magnetic modes $p \sim g^2T$ cause the
well-known perturbative breakdown \cite{Linde:1980ts}. 
However, perturbation theory restricted to the ``hard'' scale $p \sim 2\pi T$
can be treated with conventional weak-coupling methods, 
while the soft and ultra-soft scales are only accessible through
improved analytic methods or non-perturbatively via lattice simulations, as is
especially the case for the ultra-soft $g^2T$ scale. 
Here $p$ denotes the characteristic momentum scale, $g$ the gauge coupling 
and $T$ the temperature. 
The infrared problems which cause the breakdown of perturbation theory 
can be isolated into a three-dimensional (3D) effective field theory 
called magnetostatic QCD (MQCD) and studied non-perturbatively with 
lattice simulations. 
Before computing various quantities in this framework a
number of perturbative ``matching'' computations are 
necessary \cite{Kajantie:1995dw,Braaten:1995jr},
in order to relate the parameters of the effective theory
with those of thermal QCD.

The plan of this paper is the following. 
In Section~\ref{se:gEmE} we review the most important facts 
of the dimensionally reduced effective field theory framework 
and show how to systematically determine
the effective gauge coupling $\gE$ and screening mass $\mE$. 
In Section~\ref{se:reduction} we explain some technical details 
about the integral reduction step, 
while in Section~\ref{se:structure} we discuss the structure of
the explicit result for 
the one-, two-, and three-loop corrections, whose rather lengthy 
coefficients are detailed in the Appendix.
Section \ref{se:mintegrals} contains the evaluation of a new class of master 
sum-integrals that appear in our result.
We finally discuss possible applications of our results
in \se\ref{se:applications},
before we conclude in Section \ref{se:conclusions}. 

%
\section{Effective gauge coupling and screening mass}
\la{se:gEmE}

We consider QCD at finite temperature with the gauge group SU($\Nc$)
and $\Nf$ massless flavors of quarks. 
Before gauge fixing, the bare Euclidean Lagrangian in dimensional 
regularization reads
\begin{align}
S_{QCD} &= \int_0^{1/T}\!\!\!\text{d}\tau\int \text{d}^{d}x\,\mathcal{L}_{QCD}\,,\\
\la{eq:sqcd}
\mathcal{L}_{QCD} &= \frac{1}{4}F_{\mu\nu}^{a}F_{\mu\nu}^{a} + \bar{\psi}\gamma_{\mu}D_{\mu}\psi,
\end{align}
where $T$ is the temperature;
$d = 3 - 2\e$ denotes the number of spatial dimensions,
such that Greek indices run as $\mu,\nu = 0,\dots,d$;
$F_{\mu\nu}^{a} = \partial_{\mu}A_{\nu}^{a} - \partial_{\nu}A_{\mu}^{a} 
+ gf^{abc}A_{\mu}^{b}A_{\nu}^{c}$ and
$D_{\mu} = \mathbbm{1}\partial_{\mu} - igA_{\mu}^{a}T^{a}$,
where the
$T^{a}$ are hermitian generators of SU($\Nc$) with normalization
$\mathrm{Tr}[T^{a}T^{b}] = \delta^{ab}/2$;
we use hermitian Dirac matrices
$\gamma_{\mu}^{\dagger} = \gamma_{\mu}$, 
$\{\gamma_{\mu},\gamma_{\nu}\} = 2\delta_{\mu\nu}$;
$g$ is the bare gauge coupling; 
and $\psi$ carries Dirac, color, and flavor indices. 
For the group theory factors, we us the standard symbols 
$\CA = \Nc$, $\CF = (\Nc^2 - 1)/(2\Nc)$. 

At sufficiently high temperatures, the long-distance physics 
of \eq\nr{eq:sqcd} can be described by a simpler, dimensionally 
reduced effective field 
theory \cite{Ginsparg:1980ef,Appelquist:1981vg,Kajantie:1995dw,Braaten:1995jr}:
\begin{align}
        S_{\mathrm{EQCD}} &= \int \mathrm{d}^{d}x\,\mathcal{L}_{\mathrm{EQCD}}\,,\\
        \la{eq:seqcd}
        \mathcal{L}_{\mathrm{EQCD}} &= \frac{1}{4}F_{ij}^{a}F_{ij}^{a} + \mathrm{Tr}[D_{i},B_0]^2 + \mE^2\mathrm{Tr}[B_0^2] + \lE^{(1)}\mathrm{Tr}[B_0^2]^2 + \lE^{(2)}\mathrm{Tr}[B_0^4]
        +\dots \;,
\end{align}
where $i = 1,\dots,d,\,F_{ij}^{a} = \partial_{i}B_{j}^{a}
- \partial_{i}B_{j}^{a} + \gE f^{abc}B_{i}^{b}B_{j}^{c}$ and $D_{i}
= \partial_{i} - i\gE B_{i}$. 
The electrostatic gauge fields $B_0^{a}$ and magnetostatic
gauge fields $B_{i}^{a}$ appearing in the theory above can be related 
(up to normalization) to the zero modes of $A_{\mu}^{a}$ 
of thermal QCD in \eq\nr{eq:sqcd}.

The effective parameters in \eq\nr{eq:seqcd}, which we are ultimately
interested in, can be obtained by matching. 
This means, we require the same result on the QCD and EQCD side within the
domain of validity. 
A convenient way to perform the matching computation is to use a 
strict perturbation expansion in $g^2$. 
On both sides, the expansion is afflicted with infrared divergences. 
These divergences are screened
by plasma effects and can be taken into account (at least for electrostatic
gluons) by resumming an infinite set of diagrams. 
Screening of magnetostatic gluons is a completely
non-perturbative effect. 
For the matching computation, it is not necessary to worry about the infrared divergences because the matching parameters are only sensitive to the effects of large momenta. 
All infrared divergences which occur can be removed by choosing a convenient
infrared cutoff. 
It is essential to choose the same infrared cutoff in both theories.

%
\subsection{Relation for $\mE^2$}
\la{se:mE}

In order to establish a relation between the parameters of the
theories Eqs.~\nr{eq:sqcd},\nr{eq:seqcd}, consider the electric
screening mass $\mel$, 
defined in the full theory\footnote{In the presence of an infrared cut-off; otherwise, a
non-perturbative definition is needed.} by the pole of the 
static $A_0^{a}$ propagator, 
\begin{align}
\left. 0 = p^{2} + \Pi_{00}(p^{2}) 
\right|_{p_0=0,\, \vec{p}^{2} = -\mel^{2}}\,.
\la{eq:pifull}
\end{align}
On the effective theory side, the electric screening mass is, 
equivalently, defined as the pole of the 3d adjoint scalar 
$B_0$ propagator,
\begin{align}
\left. 0 = \vec p^{2} + \mE^{2} + \PiEQCD(\vec p^{2})
\right|_{\vec p^{2} = -\mel^{2}} \,,
\la{eq:piE}
\end{align}
where $\PiEQCD$ denotes the $B_0$ self-energy on EQCD side.

Noting that the self-energies start at one-loop order, the leading-order
solutions for $\mel^2$ will be suppressed by the respective
coupling parameters, such that $p^2$ is to be regarded perturbatively 
small, hence allowing for a Taylor expansion of the ``on-shell'' self-energies
around zero. For Eq.~\nr{eq:pifull}, one needs
(let us write $\PiE\equiv\Pi_{00}$ from now on)
\begin{align}
\PiE(-\mel^2) &=  \PiE(0) -\mel^2\PiE'(0) + \dots\nn
&= \sum_{n=1}^{\infty}g^{2n}\, \Pi_{{\rm E}n}(0)
-\mel^2\sum_{n=1}^{\infty}g^{2n}\,\Pi'_{{\rm E}n}(0) + \dots\,,
\la{eq:taylorpi}
\end{align}
where in a second step we have introduced the $n$\/-loop self-energy 
coefficients $\Pi_{{\rm E}n}$. From \eqs\nr{eq:pifull} 
and \nr{eq:taylorpi}, we can 
express the electric screening 
mass $\mel^2$ in terms of Taylor coefficients up to next-to-next to leading
order (NNLO)
\begin{align}
\mel^{2} &= g^{2}\PiEa + g^{4}\lk\PiEb
  - \PiEap\PiEa\rk + g^{6}\big[\PiEc 
  - \PiEap\PiEb\;- \nn
&\quad- \PiEbp\PiEa + \PiEapp\(\PiEa\)^2 
+ \PiEa\(\PiEap\)^2\big] + \order{g^{8}}\,.
\la{eq:screenme}
\end{align}
Diagrams contributing to the various orders of $\Pi$ are depicted 
in \fig\ref{fig:2pt}.

\begin{figure}
\centering
\begin{eqnarray*}
%
%
\TopoS(\Legl) & \equiv &
\sy{}12 \TopoSB(\Legl,\Agl,\Agl)
\sm{-1} \TopoSB(\Legl,\Agh,\Agh)
\sm{-1} \TopoSB(\Legl,\Aqu,\Aqu)
\sy+12 \TopoST(\Legl,\Agl)
\sm{-1} \TopoST(\Legl,\Agh) \;,
\\[0ex]
&& \nn[0ex]
%
%
\ToptS(\Legl) &\hspace{-2mm} \equiv \hspace{-1mm}& 
\sy{}12 \ToptSM(\Legl,\Agl,\Agl,\Agl,\Agl,\Lgl)
\sm{-1} \ToptSM(\Legl,\Agh,\Agl,\Agl,\Agh,\Lgh)
\sm{-1} \ToptSM(\Legl,\Agl,\Agh,\Agh,\Agl,\Lagh)
\sm{-1} \ToptSM(\Legl,\Agh,\Agh,\Agh,\Agh,\Lgl)
\sm{-1} \ToptSM(\Legl,\Aqu,\Agl,\Agl,\Aqu,\Lqu)
\sm{-1} \ToptSM(\Legl,\Agl,\Aqu,\Aqu,\Agl,\Laqu)
\sm{-1} \ToptSM(\Legl,\Aqu,\Aqu,\Aqu,\Aqu,\Lgl) \nn[1ex]&&{} \hspace*{-0.3cm}
\sy+12 \ToptSAl(\Legl,\Agl,\Agl,\Agl,\Agl)
\sy+12 \ToptSAr(\Legl,\Agl,\Agl,\Agl,\Agl) 
\sm{-1} \ToptSAl(\Legl,\Agl,\Agh,\Agl,\Agh)
\sm{-1} \ToptSAr(\Legl,\Agh,\Agl,\Agl,\Agh) 
\sm{-2} \ToptSAl(\Legl,\Agh,\Agh,\Agh,\Agl)
\sm{-2} \ToptSAr(\Legl,\Agh,\Agh,\Agh,\Agl)
\sy+14 \ToptSE(\Legl,\Agl,\Agl,\Agl,\Agl)   \nn[2ex]&&{} \hspace*{-0.3cm}
\sy+16 \ToptSS(\Legl,\Agl,\Agl,\Lgl) 
\sm{-1} \ToptSS(\Legl,\Agh,\Agh,\Lgl)
%
%
\sy+12 \ToptSBB(\Legl,\Agl,\Agl,\Agl,\Agl)
\sm{-1} \ToptSBB(\Legl,\Agl,\Agl,\Agh,\Agh)
\sm{-2} \ToptSBB(\Legl,\Agh,\Agh,\Agl,\Aagh)
\sm{-1} \ToptSBB(\Legl,\Agl,\Agl,\Aqu,\Aqu)
\sm{-2} \ToptSBB(\Legl,\Aqu,\Aqu,\Agl,\Aaqu) \nn[2ex]&&{} \hspace*{-0.3cm}
\sy+12 \ToptSBT(\Legl,\Agl,\Agl,\Agl)
\sy+14 \ToptSTB(\Legl,\Agl,\Agl,\Agl)
\sy-12 \ToptSTB(\Legl,\Agl,\Agh,\Agh) 
\sm{-1}\ToptSTB(\Legl,\Agh,\Agl,\Aagh)
\sy-12 \ToptSTB(\Legl,\Agl,\Aqu,\Aqu) 
\sy+14 \ToptSTT(\Legl,\Agl,\Agl)  \;,
\\[0ex]
&& \nn[0ex]
%
%
\ToprS(\Legl) & \equiv &
\sm{1} \ToprSMi(\Legl,\Agl,\Agl,\Agl,\Agl,\Lgl,\Lgl,\Lgl)
\sm{+1} \ToprSMiV(\Legl,\Agl,\Agl,\Agl,\Agl,\Lgl,\Lgl)
\sy+14 \ToprSMiNp(\Legl,\Agl,\Agl,\Agl,\Agl,\Lgl,\Lgl,\Lgl)
\sy+14 \ToprSMiB(\Legl,\Agl,\Agl,\Agl,\Agl,\Agl)
\sy+14 \ToprSMiBB(\Legl,\Agl,\Agl,\Agl,\Agl,\Lgl,\Lgl)
\sy+12 \ToprSMiBT(\Legl,\Agl,\Agl,\Agl,\Lgl,\Lgl,\Lgl)
\sm{+\;\;\;441\;\text{diags}}\;.
\end{eqnarray*}
\caption{The 1-loop, 2-loop and some 3-loop self-energy diagrams 
in the background field gauge. 
Wavy lines represent gauge fields, dotted lines ghosts,  and solid lines fermions.}
\la{fig:2pt}
\end{figure}

To complete the matching computation for $\mel^2$, we have to compute
$\PiEQCD$ on the EQCD side in a strict perturbative expansion. 
Again treating the ``on-shell'' momentum $\vec p^2$ (as well as the
tree-level mass $\mE^2$) as perturbatively small,
due to the fact that the only scale in $\PiEQCD(\vec p^{2})$ is $\vec p^{2}$,
after Taylor expansion 
the dimensionally regularized integrals (being scale-free) 
vanish identically\footnote{Note that this is not the case for the 
coefficients of Eq.~\nr{eq:taylorpi}, since those are vacuum
sum-integrals in the full theory 
and hence know about the temperature scale $T$.}. From \eq\nr{eq:piE} 
it hence follows that
\begin{align}
         \mE^{2} = \mel^{2}\,.
\end{align}

%
\subsection{Relation for $\gE^2$}
\la{se:gE}

In order to relate the effective 3d gauge coupling $\gE^2$ 
to the parameters of the full theory,
we can choose whether to go through a 3-point or a 4-point 
function, in addition to a 2-point function. 
However, it is further possible to simplify this task to a single 2-point
calculation using the background field gauge method (see e.g. 
Ref.~\cite{Abbott:1980hw}). 
Let us give the main argument here, closely following Ref.~\cite{Laine:2005ai}.

The effective Lagrangian \eq\nr{eq:seqcd} follows from integrating out 
the hard ($p\sim T$) scales which, symbolically, produces an expression
of the form
\begin{align}
\la{eq:efflangsym}
\mathcal{L}_{\text{eff}} \sim c_2(\partial B)^2 
+c_3 g (\partial B) B^2 +c_4g^2 B^4 + \dots\;,
\end{align}
where $B$ denotes the background field potential and the coefficients 
$c_{i} = 1 +\order{g^{2}}$. Redefining now the effective field as
$B_{\text{eff}}^{2} \equiv c_{2}B^{2}$, from
$\mathcal{L}_{\text{eff}} \sim (\partial B_{\text{eff}})^2 
+c_3c_2^{-3/2}g(\partial B_{\text{eff}})B_{\text{eff}}^2 
+ c_4c_2^{-2}g^2B_{\text{eff}}^4 + \dots$
we can read off the effective gauge coupling 
(considering the gauge invariant structure $F^2$)
$g_{\text{eff}} = c_{3}c_{2}^{-3/2}g = c_{4}^{1/2}c_{2}^{-1}g$.
Furthermore, since the effective action is gauge invariant with
respect to both $B_{\text{eff}}$ as well as $B$ \cite{Abbott:1980hw},
we have $c_{2} = c_{3} = c_{4}$.
Finally transforming to 3d notation, scaling the fields 
$B\!\rightarrow\!T^{1/2}B^2$
and comparing $\int_0^{1/T}\!\!{\rm d}\tau\,{\cal L}_{\rm QCD}$
with ${\cal L}_{\rm EQCD}$, it follows that 
\begin{align}
\la{eq:ge2}
\gE &= T^{1/2}\,c_{2}^{-1/2}\,g\;.
\end{align}

Now we proceed in the same way with the effective gauge coupling $\gE$ as for
the screening mass $\mE$. From \eq\nr{eq:ge2} we thus obtain
\begin{multline} 
\la{eq:taylorge2}
 \gE^{2} = T\lb g^{2} - g^{4}\PiTa
 + g^{6}\lk\(\PiTap\)^{2} - \PiTbp\rk + \right.\\
 \left.+ g^{8}\lk 2\,\PiTap\PiTbp -
 \(\PiTap\)^{3} - \PiTcp\rk 
 + \order{g^{10}} \rb\,,
\end{multline}
where $\PiT$ denotes the transverse part of the (spatial 
part of the) self-energy
\begin{align}
\la{eq:selfstruct}
\Pi_{ij}(\vec{p}) \equiv 
\(\delta_{ij} - \frac{p_{i}p_{j}}{\vec{p}^{2}}\)\PiT(\vec{p}^{2}) 
+ \frac{p_{i}p_{j}}{\vec{p}^{2}}\Pi_{\text{L}}(\vec{p}^{2})\,.
\end{align}
To understand the split-up of $\Pi_{\mu\nu}$ in more detail, note that we can
choose the external momentum $p$ purely spatial, 
$p = (0,\vec{p})$, while the rest frame of the heat bath is time-like, with 
Euclidean four-velocity $u = (1,0)$, such that $u\cdot u = 1, u \cdot p = 0$. 
In this case $\Pi_{\mu\nu}$ has three independent components ($\Pi_{0i}$,
$\Pi_{i0}$ vanish identically). 
The loop corrections to 
the spatially longitudinal part $\Pi_\rmi{L}$ also vanish 
(which we will however explicitly check in our computations), 
such that only two non-trivial functions, $\PiE$ and $\PiT$, remain
(recall $\PiE=\Pi_{00}$).

Noting that the class of background field gauges 
still allows for a general gauge parameter $\xi$ 
(we denote $(\xi)_\mathrm{here} = 1 - (\xi)_\mathrm{standard}$),
we use the gauge field propagator
\begin{align}
\la{eq:propag}
D_{\mu\nu}^{ab}(q) = 
\delta^{ab}\lk\frac{\delta_{\mu\nu}}{q^2} - 
\xi\frac{q_{\mu}q_{\nu}}{(q^2)^2}\rk
\end{align}
and verify gauge parameter cancellation in the end of our computations.

%
\section{The reduction}
\la{se:reduction}

After the Taylor expansion and decoupling of scalar products with external 
momentum, all integrals that contribute to the self-energies up to 
three-loop order that are needed for Eqs.~\nr{eq:screenme} 
and \nr{eq:taylorge2} can be written as
\begin{align}
\la{eq:sc3l}
\II_{a,b,c,d,e,f;\,c_1,c_2,c_3}^{\alpha,\beta,\gamma} \equiv 
\sumint{P_1 P_2 P_3}
\frac{(P_1)_0^{\alpha}\,(P_2)_0^{\beta}\,(P_3)_0^{\gamma}}
{[P_1^{2}]^{a}\,[P_2^{2}]^{b}\,[P_3^{2}]^{c}\,
[(P_1-P_2)^{2}]^{d}\,[(P_1-P_3)^{2}]^{e}\,[(P_2-P_3)^{2}]^{f}}\,,
\end{align}
where $P_i^2 = (P_i)_0^2 + \vec{p}_i^{2} = [(2n_i+c_i)\pi T]^2+\vec{p}_i^{2}$ 
for $i\in\{1,2,3\}$ are bosonic (fermionic) loop momenta for $c_i=0$ ($1$). 
The sum-integral symbol
in \eq\nr{eq:sc3l} is a shorthand for
\begin{align}
\sumint{P} \rightarrow 
\mu^{2\e}T\sum_{P_0}\int\frac{\mathrm{d}^{d}p}{(2\pi)^{d}}\,,
\end{align}
where $\mu$ is the minimal subtraction ($\text{MS}$) 
scheme scale parameter, and we take $d=3-2\e$.

An essential part of this work deals with the reduction of integrals of the
type in \eq\nr{eq:sc3l} to a small set of master integrals. 
We use the well-known integration by parts (IBP) identities
and identities following from exchanges of integration variables. 
Both are implemented in a Laporta algorithm 
\cite{Laporta:2001dd} using {\tt FORM} \cite{Vermaseren:2000nd}.
Compared to the well-established Laporta-type algorithms for zero-temperature
reductions, one of the main differences here is that the IBP relations
act only within the continuum (spatial) part of our sum-integrals. 
Another important difference is that in general, linear shifts or 
exchanges of integration momenta can cause a flip of bosonic and 
fermionic signature of the loop momenta, such that extra care must
be taken for topology mapping. 
A precursor of this reduction algorithm had already been tested
in Ref.~\cite{Laine:2005ai}.

\begin{figure}
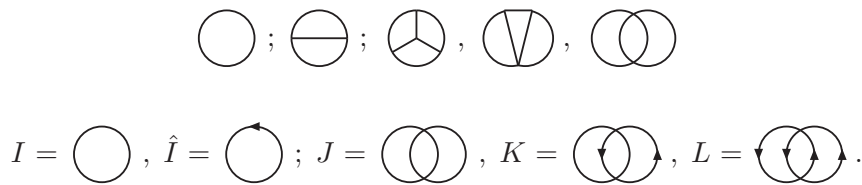

\centering
\begin{align*}
\TopoVR(\Aqq)\,;\;
&\ToptVS(\Aqq,\Aqq,\Lqq)\,;\;
\ToprVM(\Aqq,\Aqq,\Aqq,\Lqq,\Lqq,\Lqq)\,,\;
\ToprVV(\Aqq,\Aqq,\Lqq,\Lqq,\Lqq)\,,\;
\ToprVB(\Aqq,\Aqq,\Aqq,\Aqq)
\end{align*}
\begin{align*}
\Ib{}{} = \TopoVR(\Aqq)\,,\;
\If{}{} = \TopoVR(\Aqu)\,;\;
\JJ{}{} = \ToprVB(\Aqq,\Aqq,\Aqq,\Aqq)\,,\;
\KK{}{} = \ToprVB(\Aqu,\Aqu,\Aqq,\Aqq)\,,\;
\LL{}{} = \ToprVB(\Aqu,\Aqu,\Aqu,\Aqu)\,.
\end{align*}
\caption{Top row: non-trivial vacuum topologies at 1-loop, 2-loop 
and 3-loop. Bottom row: types of bosonic and fermionic master integrals. 
Lines (arrow-lines) corresponds to bosonic (fermionic) propagators,
respectively.}
\la{fig:0pt}
\end{figure}

The main difference between the outcome of the
1-loop and 2-loop calculation on the one side 
and the 3-loop correction on the other side is that the 
former ones are expressible in terms of 1-loop 
tadpole sum-integrals which are known explicitly, see App.~\ref{app:1loop}. 
This is no longer the case at 3-loop order. 
The mercedes- and spectacles topology shown on the 
first line of \fig\ref{fig:0pt} can be expressed in terms of 
basketball-type sum-integrals as well as products of 1-loop tadpoles.

%
\section{Structure of the result}
\la{se:structure}

After reduction, we can express all quantities as a sum of 1- and 3-loop 
master integrals (there are no master integrals at 2-loop order, 
see \cite{Braaten:1995jr,Schroder:2008ex}) of the generic types depicted
on the second row of Fig.~\ref{fig:0pt},
the structure being
\begin{align}
&\Pi_3 = \sum_i a_i\,A_i + \sum_j b_j\,B_j \;,\\
&\mbox{where}\;\; A_i = \II \cdot \II \cdot \II \quad\mbox{with}\quad 
 \II\in\lb\Ib{m}{n},\If{m}{n}\rb\\
&\mbox{and}\;\; B_j = {\rm basketball} \in \lb \JJ{}{},\KK{}{},\LL{}{}\rb \;.
\end{align}
A detailed version is given in the Appendix, 
cf.\ \eqs\nr{eq:alphas} and \nr{eq:betas}.

We have performed a number of cross-checks to confirm the validity of our results: 
the longitudinal parts of the self-energy vanish identically 
\begin{align}
\Pi_{\mathrm{L3}} &= \Pi'_{\mathrm{L3}} = 0 \quad\text{for}\quad \xi^0,\dots, \xi^6\,,
\end{align}
and the specific combinations of (bare) self-energy coefficients that
build up $\mE^2$ (cf. \eq\nr{eq:screenme}) and $\gE^2$ (cf. \eq\nr{eq:taylorge2}) 
are gauge-parameter independent up to three-loop order.

The one-, and two-loop
calculations have already been performed in Ref.~\cite{Laine:2005ai} which we use
as another serious cross-check of our independent calculation. We obtain 
full agreement when comparing our Eqs.~\nr{eq:C1}--\nr{eq:C4} and 
Eqs.~\nr{eq:C7}--\nr{eq:C10} with that reference.

There is considerable experience of how to calculate the genuine 3-loop 
integrals $B_j$ up to the constant term (which can typically only be 
represented in terms of two-dimensional parameter integrals and evaluated 
numerically), see \cite{Arnold:1994ps,Gynther:2007bw,Andersen:2008bz}. 
In Section~\ref{se:mintegrals}, we add to this available knowledge a specific 
class of 3-loop (basketball-type) sum-integrals which appear in
our reduced expressions \eqs\nr{eq:alphas} and \nr{eq:betas}.

It turns out, however, that most of the pre-factors $b_j$ are singular when
expanded around $d=3-2\e$ dimensions. 
Hence, we need to expand the integrals
$B_j$ beyond their constant term (in fact, to $\order{\e}$ for $\PiE$ and to
$\order{\e^2}$ for $\PiT'$). 
As the conventional techniques for computing
these basketball-type integrals rely on a careful subtraction of
sub-divergences on a case-by-case basis, it appears quite difficult to
extend the known techniques in order to evaluate higher terms in the epsilon
expansion.
 
To make progress, it might be advantageous to perform a change of basis, see
e.g. \cite{Chetyrkin:2006dh}, in order to avoid or at least reduce the
number of divergent pre-factors. 
Due to the large number of integrals contained
in our reduction tables, an algorithmic approach trying out all possible
different combinations of basis elements might be somewhat involved, but
certainly possible.

%
\section{Evaluation of classes of master sum-integrals}
\la{se:mintegrals}

After the successful reduction step, a number of non-trivial three-loop
master sum-integrals will have to be evaluated. 
Noting that all bosonic and fermionic one-loop sum-integrals 
$I_m^n$ and $\hat I_m^n$ that appear in \eqs\nr{eq:C1}--\nr{eq:C10} as
well as in \eqs\nr{eq:alphas}, \nr{eq:betas} are known analytically
(see App.~\ref{app:1loop}), and noting that furthermore all 2-loop structures
have been reduced to products of 1-loop integrals,
let us tackle 
the first non-trivial sub-class of master integrals, the bosonic basketball
\begin{align}
\la{eq:bmaster1}
B_{N,M} \equiv 
\II_{N,1,0,0,1,1;\,0,0,0}^{M,0,0}
= \sumint{PQR}\frac{Q_0^{M}}{[Q^2]^{N}\,(P-Q)^2\,R^2\,(P-R)^2}\,,
\end{align}
with $N,M\geq2$. 
After a careful subtraction of all UV and IR divergences 
(for more details 
see \cite{Arnold:1994ps,Gynther:2007bw,Moller:2010xw,Moeller:Thesis:2009}) 
we can write \eq\nr{eq:bmaster1} as 
\begin{align}
\la{eq:bmaster2}
B_{N,M} &= \beta\lk A(N,\e,1)\delta_{M,0} + \bar\beta \Ib{N-1+2\e}{M} 
  + \Ib{1}{0}\,\Ib{N+\e}{M}\rk + B_{N,M}^{IV}
+\nn&+
  2 \Ib{1}{0}\,S(N,1,1;M,0) + \sumint{PQ}
  \frac{\Delta\Pi(P)\delta_{P_0}\delta_{Q_0}}{[Q^2]^{N}\,(P-Q)^2}\,
  \delta_{M,0} + B_{N}^{II}\delta_{M,0}+ B_{N,M}^{I}\,,
\end{align}
where $\beta\equiv G(1,1,d+1)$ stands for the 4d massless 
1-loop bubble and $\bar\beta \equiv G(\frac{3-d}{2},1,d+1)$ 
is the 4d 1-loop propagator, where the function $G$ reads
($s_{12}\equiv s_1+s_2$ etc.)
\begin{align}
G(s_1,s_2,d) &\equiv (p^2)^{s_{12}-\frac{d}{2}}
\int_q\frac{1}{[q^2]^{s_1}[(q-p)^2]^{s_2}}
= \frac{\Gamma{(\frac{d}{2}-s_1})\Gamma{(\frac{d}{2}-s_{2})
\Gamma{(s_{12}-\frac{d}{2}})}}{(4\pi)^{d/2}\Gamma{(s_1)}
\Gamma{(s_2)}\Gamma{(d-s_{12})}}\,,
\end{align}
and $S$ stands for the two-loop tadpole at finite-temperature, 
\begin{align}
\la{eq:Sdef}
S(s_1,s_2,s_3;a_1,a_2) &\equiv 
\sumint{PQ} \frac{|Q_0|^{a_1}\,|P_0|^{a_2}}
{[P^2]^{s_1}\,[Q^2]^{s_2}\,[(P-Q)^2]^{s_3}} \nn
&= \sum_i \Ib{i}{0}\,\Ib{s_{123}-a_{12}/2-i}{0}\,e_i(s_1,s_2,s_3,a_1,a_2,d)\,,
\end{align}
where the coefficients $e_i$ follow from IBP relations (for an example, 
see \eq\nr{eq:ei} below).
Furthermore, the abbreviation $A(s_1,s_2,s_3)$ stands for a specific 2-loop tadpole
\begin{align}
A(s_1,s_2,s_3) &\equiv \sumint{PQ}
   \frac{\delta_{Q_0}}{[Q^2]^{s_1}[P^2]^{s_2}[(P-Q)^2]^{s_3}} 
   = \frac{2T^2\zeta{(2s_{123}-2d)}}{(2\pi T)^{2s_{123} -2d}}
  \,N(s_1,s_2,s_3)\,,\\
N(s_1,s_2,s_3) &\equiv \int_{pq}
   \frac{1}{[p^2+1]^{s_1}[q^2+1]^{s_2}[(p-q)^2]^{s_3}} 
\nn&=
   \frac{\Gamma{(s_{13}-\frac{d}{2})}\Gamma{(s_{23}-\frac{d}{2})}
   \Gamma{(\frac{d}{2}-s_3)}\Gamma{(s_{123}-d)}}
   {(4\pi)^d\Gamma{(s_1)}\Gamma{(s_2)}\Gamma{(d/2)}\Gamma{(s_{1233}-d)}}\,.
\end{align}
In \eq\nr{eq:bmaster2} we make use of the one-loop subtracted quantities
\begin{align}
\Delta\Pi(P) &= \sumint{R} \frac1{R^2\,(R-P)^2}-\frac{\beta}{[P^2]^\e}
-\frac{2I_1}{P^2}\;, \\
\Delta\tilde\Pi(Q)&=\sumint{R} \frac1{[R^2]^\e\,(R-Q)^2}
-\frac{\bar\beta}{[Q^2]^{2\e-1}}
-\frac{2I_1}{[Q^2]^\e}\;,
\end{align}
as well as the three pieces
\begin{align}
        B_{N,M}^{I} &= \sumint{P}\sumintp{Q}\frac{\Delta\Pi(P)Q_0^{M}}
{[Q^2]^{N}(P-Q)^2}\,,\quad
        \la{eq:bn2}
        B_{N}^{II} = \sumintp{P}\sumint{Q}\frac{\Delta\Pi(P)\delta_{Q_0}}
{[Q^2]^{N}(P-Q)^2}\,,\\
        B_{N,M}^{IV} &= \beta\,\sumintp{Q}\frac{\Delta\tilde\Pi(Q)Q_0^{M}}
{[Q^2]^N} \;,
\end{align}
where the primed sums denote $\sum_n'=\sum_{n\neq0}$.
It turns out, however, that $B_{N}^{II}$ contains an additional IR divergence
which can be taken into account either by means of IBP reduction
\cite{Moller:2010xw} or by subtraction by hand
\cite{Gynther:2007bw,Moeller:Thesis:2009}, adding the appropriate zeros
(massless tadpoles which vanish in dimensional regularization). 
Performing a transformation to coordinate space in $d=3$ dimensions and 
evaluating the remaining sums give the 1d integral representations
\begin{align}
\la{eq:decJMref2}
\left.B_{N,M}^n\right|_{\e=0} &=
  \frac{T^{6-2N}\,2^{N-1}}{\Gamma(N)\,(4\pi)^{2N}}\,(2\pi T)^M
  \intr\hat B_{N,M}^n(r)\,\Delta\pi(r)\,,\\
\hat B_{N,M}^I(r) &= \sum_{i=0}^{N-2} c_{Ni}\,r^{N-3-i}\,
  \lb \li{N-2+i-M} +\coth(r)\,\li{N-1+i-M}\rb\,,\\
\hat B_N^{II}(r) &= -\sum_{n=0}^{N-2}\sum_{i=0}^{N-2+n}
  \frac{\Gamma(N)}{\Gamma(N+n)}\,\frac{a_{N,n}}{2^n}\,c_{N+n,i}\,
  r^{N-2+n-i}\,\li{N-2+i-n}\,,\\
\hat B_{N,M}^{IV}(r) &= \sum_{i=0}^{N-2} c_{Ni}\,r^{N-3-i}
  \lb \frac{N\!-\!1\!-\!i}{2r}\,\li{N-1+i-M}-\frac12\,\li{N-2+i-M} \rb\,,
\end{align}
with $\Delta\pi(r)\equiv\coth(r)-\frac1r-\frac{r}3$ and 
where the $c_{N,i}$ are Fourier coefficients given by
\begin{align}
        \sqrt{\frac{2m}{\pi}}e^{m}K_{3/2-s}(m) = \sum_{n=0}^{\mathrm{max}(s-2,1-s)}\frac{c_{s,n}}{m^n}
\end{align}
and $a_{N,n}$ can be obtained by IBP reduction of the inner sum-integral of \eq\nr{eq:bn2}, see \cite{Moller:2010xw}.

Putting all ingredients together for the special case $B_{3,2}$ (which is 
needed for $\mE^2$, being the coefficient of $\alpha_4$ in \eq\nr{eq:alphas}), 
evaluating the finite pieces numerically,
\begin{align}
\left.B_{3,2}^{I}\right|_{\e=0} &= \frac{T^2}{2(4\pi)^4} \intr \Delta\pi(r) 
   \lb \li{-1}\!+\!\(\coth(r)\!+\!\frac1r\)\!\li{0}\!+\!\frac{\coth(r)}{r}\,\li{1}\rb
\nn&\approx -\frac{T^2}{2\,(4\pi)^4}\,\times\, 0.029779678110507967168(1)\,,\\
\left.B_{3,2}^{IV}\right|_{\e=0} &= \frac{T^2}{2\,(4\pi)^4}\, \intr \Delta\pi(r)
   \lb -\frac12\,\li{-1}+\frac1{2r}\,\li{0}+\frac1{2r^2}\,\li{1}\rb
\nn&\approx -\frac{T^2}{2\,(4\pi)^4}\,\times\, 0.0020065925001817061293(1)\,,
\end{align}
and using (from IBP, see \eq\nr{eq:Sdef})
\begin{align}
\la{eq:ei}
e_{i}(3,1,1,2,0,d) = \frac{(d-4)^2}{(d-2)(d-5)(d-7)}\,\delta_{i,2}\,,
\end{align}
we obtain as final result for this new master integral 
(with $Z_1'\equiv\zeta'(-1)/\zeta(-1)$)
\begin{align}
\la{eq:finalB32}
B_{3,2} &= \frac{T^2\,(4\pi T^2)^{-3\e}}{32\,(4\pi)^4\,\e^2}
   \lk 1 + \(\frac{41}6+\gammaE+2Z_1'\)\e 
   + 70.32026114816592109(1)\,\e^2 +\order{\e^3}\rk\,.
\end{align}
For an important cross-check of this result, see App.~\ref{app:check}.

%
\section{Applications}
\la{se:applications}

To emphasize the necessity to pursue the matching computations as
outlined in this note, let us briefly discuss two applications
that would become relevant once full results are available.

The first immediate application involves the Debye screening mass
$\mE^2$ of \se\ref{se:mE} and concerns higher-order
perturbative contributions to basic thermodynamic observables,
such as the pressure of hot QCD.
In fact, once the quantity $\PiEc$ of \eq\nr{eq:alphas} has been 
fully determined, the mass term of EQCD (cf.\ \eq\nr{eq:seqcd})
is available at NNLO, $\mE^2\sim g^2T^2[1+g^2+g^4+\order{g^6}]$,
where $g$ is the dimensionless gauge coupling of full QCD.
Now, in the context of the effective theory setup for hot QCD,
it turns out that the lowest-order EQCD contribution to the full 
pressure, coming from the quadratic part of $\mathcal{L}_{\mathrm{EQCD}}$,
enters as $\sim T\mE^3$ \cite{Braaten:1995jr},
which translates to $T^4 g^3[1+g^2+g^4+\order{g^6}]$,
such that our 3-loop coefficient contributes to $\order{g^7}$
in the QCD pressure. 
According to the systematics of effective theory, due
to the fact that there are typically large logarithms,
a systematic $g^6$ evaluation of the pressure
(almost completely known at present, only missing a 
well-defined perturbative 4-loop 
computation \cite{Braaten:1995jr,Kajantie:2000iz})
has actually been coined {\em physical leading order}, 
since it is the first order where all three
physical scales (hard/soft/ultra-soft) have contributed.
In this respect, the $\order{g^7}$ term would simply be 
next-to-leading order, and allow for a first 
serious investigation of convergence properties.

Leaving the incomplete $\order{g^6}$ (for which there exist
numerical estimates, however, from comparisons with lattice 
data, see e.g.\ \cite{Laine:2006cp})
aside for the moment, 
there are other sources of $\order{g^7}$ contributions, of course:
from the the MQCD pressure plus NLO matching of the 3d MQCD gauge
coupling $\gM^2$; 
from the terms proportional to the quartic coupling $\lE$ in 
the 3-loop EQCD pressure;
from the 5-loop EQCD pressure (at $\lE=0$), which entails one of the
conceptually simplest (3d, super-renormalizable, massive, vacuum-diagram) 
computations at the 5-loop level, for which techniques are
presently developed by several groups;
and from the leading terms of some higher-order operators
in the EQCD Lagrangian, denoted by dots in \eq\nr{eq:seqcd},
but classified in \cite{Chapman:1994vk}.
All but the last two of these additional $g^7$ contributions 
are already known.

A second immediate application involves the 3d EQCD gauge coupling 
$\gE^2$ of \se\ref{se:gE}
and concerns precision-tests of the dimensional reduction 
setup, such as for the spatial string tension $\sigma_s$,
which parameterizes the large-area behavior of rectangular spatial 
Wilson loops. 
As has been demonstrated in Ref.~\cite{Laine:2005ai},
it can be systematically determined, as a function of the temperature $T$, 
in the dimensionally reduced effective theory setup, and then
compared to non-perturbative 4d lattice measurements.
It turned out that the NLO result for $\gE^2$ as obtained in 
\cite{Laine:2005ai} represents a considerable improvement
over a 1-loop comparison -- giving a sizable correction factor
as well as a first estimate of (renormalization) scale 
dependence -- while leaving room for NNLO effects,
for which our 3-loop result for $\PiTcp$ of \eq\nr{eq:betas} 
is the last missing building block.

%
\section{Conclusions}
\la{se:conclusions}

We have successfully reduced the NNLO contributions to the 
matching parameters $\mE^2$ and $\gE^2$ to a sum of scalar
sum-integrals. 
These matching parameters play an important role in higher-order
evaluations of basic thermodynamic observables and in precision-tests
of the dimensional reduction setup respectively, 
and hence are needed with high accuracy.
Our result passes the non-trivial checks of transversality 
as well as gauge-parameter independence.

In a next step, a number of master integrals have to be evaluated.
Although we managed to map all of them to the relatively simple
class of basketball-type ones, the somewhat large number of masters 
which we need demand a semi-automated evaluation strategy, which
still has to be developed. 
As a first and encouraging step towards this goal, we have demonstrated 
a systematic method to evaluate a certain class
of such basketball-type sum-integrals. 

Once full results for the matching coefficients discussed here
become available, there are immediate applications to quantities of 
phenomenological interest, such as the pressure of hot QCD, 
or the spatial string tension, as discussed 
in \se\ref{se:applications} above. However, these concrete 
applications will have to await progress in the art of sum-integration
for now.

%
\acknowledgments

This work was supported by the Deutsche Forschungsgemeinschaft
(DFG) under contract no.~SCHR 993/2,
by the BMBF under project no.~06BI9002,
and by the Heisenberg Programme of the DFG. contract no.~SCHR~993/1.

%
\appendix

%
\section{One- and two-loop vacuum sum-integrals}
\la{app:1loop}

The one-loop bosonic tadpole is known analytically and reads
\begin{align}
\la{eq:1looptadpole}
\Ib{m}{n} \equiv \sumint{P}\frac{P_0^{n}}{(P^{2})^{m}} 
 = \frac{2 \pi^{3/2} T^4}{(2\pi T)^{2m-n}}\(\frac{\mu^2}{\pi T^2}\)^{\e}
 \frac{\Gamma\(m - \frac{3}{2} + \e\)}{\Gamma(m)}\zeta(2m - n - 3 + 2\e)\,,
\end{align}
whereas the fermionic tadpole can be related to the corresponding bosonic 
one via
\begin{align}
\la{eq:ferbosr} 
\If{m}{n} \equiv \sumint{\{P\}}\frac{P_0^{n}}{(P^{2})^{m}} 
 = (2^{2m-n-3+2\e} - 1)\Ib{m}{n}\,.
\end{align}
As mentioned above, via integration-by-parts relations
all two-loop integrals are expressible in terms of 
products of two one-loop tadpoles which means they are also available
analytically up to arbitrary order in $\e$.

%
\section{Check of new sum-integrals}
\la{app:check}

We can cross-check our new result given in \se\ref{se:mintegrals}
using IBP reduction of the V-type topology which gives
\begin{align}
V \equiv \II_{1,1,1,1,1,0;\,0,0,0}^{0,0,0}
= \frac4{3(d-3)^2}\lb 4 B_{3,2} + \frac{3d^2-24d+47}{2(d-4)}\,B_{2,0} \rb\;,
\end{align}
where V stands for the spectacles-type diagram 
given in \cite{Andersen:2008bz}:
\begin{align}
V &\equiv \sumint{PQR}
\frac1{P^2\,Q^2\,(P-Q)^2\,R^2\,(P-R)^2} \nn
&= -\frac{T^2\,(4\pi T^2 e^{\gammaE})^{-3\e}}{4\,(4\pi)^4\,\e^2}
\lb 1 + AK_1\e+AK_2\e^2+\order{\e^3}\rb \;,
\end{align}
with $AK_1=\frac43+4\gammaE+2Z_1'$, while $AK_2$ is known only numerically.
Writing the coefficients of our basketball-results, given in
\eq\nr{eq:finalB32} above as well as \eq(26) of \cite{Moller:2010xw}, 
as 
\begin{align}
B_{3,2} &= \frac{T^2\,(4\pi T^2)^{-3\e}}{32(4\pi)^4\e^2} 
\lk b_{320}+ b_{321}\e +b_{322}\e^2 + \order{\e^3}\rk\;,\\
B_{2,0} &= \frac{T^2\,(4\pi T^2)^{-3\e}}{8(4\pi)^4\e^2} 
\lk 1+ b_{21}\e +b_{22}\e^2 + \order{\e^3}\rk\;,
\end{align}
to match the leading term of $V$ it follows that the linear relations
\begin{align}
b_{320} = 1\;,\quad
b_{321} = b_{21}+4\;,\quad
\la{check3}
b_{322} = b_{22}+4b_{21}-8
\end{align}
have to be satisfied. 
Our results presented above do indeed confirm these relations,
which we take as a nice check of our generic parameterizations.
\eq\nr{check3} provides a welcome check of our numerical constants.

%
\section{Expansion coefficients up to three loops}
\la{se:taylorcoef1}

For convenience, we here repeat the one- and two-loop coefficients
that were already computed in \cite{Laine:2005ai}, adding the second 
derivatives that are needed for \eq\nr{eq:screenme}.
The one-loop coefficients up to second derivative read
\begin{align}
\PiTa &= 0\,,\la{eq:C1}\\
\PiEa &= (d-1)\Big[\CA(d-1)\Ib{1}{0} - 2 \Nf \If{1}{0}\Big]\,,\\
\PiTap &= \frac{2 \Nf}{3}  \If{2}{0} +\frac{\CA}{6} (d-25) \Ib{2}{0}\,,\\
\PiEap &= \frac{\Nf}{3}(d-1)\If{2}{0}
   - \CA\lk\frac{28 - 5d + d^{2}}{6} + (d-3)\xi\rk \Ib{2}{0}\,,\la{eq:C4}\\
\PiTapp &= \frac{\CA}{3}\lk\frac{41}{10}
   - \frac{1}{10}d + 2\,\xi - \frac{1}{4}\xi^{2}\rk \Ib{3}{0} 
   - \frac{4 \Nf}{15}\If{3}{0}\,,\\
\PiEapp &= \frac{\CA}{3}\lk\frac{23}{5} - \frac{7}{10}d
   + \frac{1}{10}d^2 + \xi\(d-3\) + \frac{\xi^2}{4}\(d-6\)\rk
   \Ib{3}{0} + \frac{\Nf}{15}(1-d)\If{3}{0}\,.
\end{align}
The two-loop coefficients up to first derivative are given by 
(see also \cite{Laine:2005ai})
\begin{align}
\PiTb &= 0\,,\la{eq:C7}\\
\PiEb &= (d-1)(d-3)\bigg\{(1 + \xi)\Big[2 \Nf \If{1}{0} -
   (d-1)\CA\,\Ib{1}{0}\Big]\CA \Ib{2}{0} +
\nn&\qquad\qquad\qquad\qquad
+ 2\,\Nf\CF\Big[\Ib{1}{0} - \If{1}{0}\Big]\If{2}{0}\bigg\}\,,\\
\PiTbp &= \frac{(d-3)(d-4)}{(d-7)(d-5)(d-2)d}
   \bigg\{(-14 - 42 d + 8 d^2)\,\CA^{2}\,\Ib{2}{0}\,\Ib{2}{0} -
\nn&- 4\Big[4 \CF + (1 -6 d + d^2)\CA\Big]\Nf \Ib{2}{0}\,\If{2}{0} 
   -\Bigg[\(\frac{d^3}{2} - 6 d^2 + \frac{39}{2}d - 6\)\CA -
\nn&- (-14 + 41 d - 12 d^2 + d^3)\CF\Bigg]\Nf \If{2}{0}\,\If{2}{0} \bigg\}
\nn&+\frac{(d-1)}{3d (d-7)}\bigg\{(144 -31d + d^2)\Big[(1-d)\CA
                \Ib{1}{0} + 2 \Nf \If{1}{0}\Big]\CA\,\Ib{3}{0} -
\nn&- 4 (d-6)(d-1)\CF\Nf\Big[\Ib{1}{0} - \If{1}{0}\Big]\If{3}{0}\bigg\}\,,\\
\PiEbp &= \frac{(d-3)}{2(d-7)(d-5)(d-2)d}\bigg\{
   \(56 + 315 d - 231 d^2 + 57d^3 - 5 d^4\)\CA^2\,\Ib{2}{0}\,\Ib{2}{0}
+\nn&+2(d-4)(d-1)\Big[\(2-5d+d^2\)\CA + 8\CF\Big]\Nf\,\Ib{2}{0}\,\If{2}{0}
+\nn&+(d-1)\Big[\(24-7d^2+d^3\)\CA -2\(28+2d-7d^2+d^3\)\CF\Big]\Nf\,
   \If{2}{0}\,\If{2}{0}\bigg\}
+\nn&+\frac{(d-3)\,\xi}{24(d-2)}\Big[
   3\(16-13d+3d^2\)\xi-4\(44-29d+7d^2-d^3\) \Big]\CA^{2}\Ib{2}{0}\,\Ib{2}{0}
-\nn&-\frac{(d-3)(d-1)}{3}\,\xi\,\CA\Nf\,\Ib{2}{0}\,\If{2}{0} 
    + \frac{(d-1)}{6(d-7)d}\bigg\{4\(6+15d-10d^2+d^3\)
\times\nn&\times\CF\Nf\Big[\If{1}{0} - \Ib{1}{0}\Big]\If{3}{0} +
                \Big[2\(-72+42d-13d^2+d^3\) + 2(d-7)d^2\xi
+\nn&+(d-7)(d-6)d\xi^2\Big]\lk(d-1)\CA\Ib{1}{0} -
                2 \Nf \If{1}{0}\rk \CA \Ib{3}{0}\bigg\}\,.
\la{eq:C10}
\end{align}

For presenting the outcome of the reduction procedure for the
three-loop contributions, which constitutes the main result of this paper, 
we denote the master integrals as in \fig\nr{fig:0pt}, i.e.
$\Ib{}{},\If{}{}$ for the 1-loop tadpoles of \eqs\nr{eq:1looptadpole} 
and \nr{eq:ferbosr}, and 
\begin{align}
\JJ{a,b,c,d,e,f}{\alpha,\beta,\gamma} &\equiv 
  \II_{a,b,c,d,e,f;\,0,0,0}^{\alpha,\beta,\gamma} \;,\\
\KK{a,b,c,d,e,f}{\alpha,\beta,\gamma} &\equiv 
  \II_{a,b,c,d,e,f;\,0,0,1}^{\alpha,\beta,\gamma} \;,\\
\LL{a,b,c,d,e,f}{\alpha,\beta,\gamma} &\equiv 
  \II_{a,b,c,d,e,f;\,1,1,0}^{\alpha,\beta,\gamma}
\end{align}
are 3-loop basketball-type integrals in a slightly more compact 
notation than \eq\nr{eq:sc3l}.
The results needed for \eqs\nr{eq:screenme} and \nr{eq:taylorge2} then read

\begin{align}
\la{eq:alphas}
\PiEc = \CA^3 \Big[ &%
\alpha_{1}\JJ{2,1,0,0,1,1}{0,0,0} + 
\alpha_{2}\JJ{2,2,0,0,1,1}{0,0,2} + 
\alpha_{3}\JJ{3,1,0,0,1,1}{0,2,0} + 
\alpha_{4}\JJ{3,1,0,0,1,1}{2,0,0} + 
\alpha_{5}\JJ{4,1,0,0,1,1}{1,3,0} + \nn&%
\alpha_{6}\JJ{5,1,0,0,1,1}{6,0,0} + 
\alpha_{7}\JJ{5,3,0,0,1,1}{6,4,0} + 
\alpha_{8}\JJ{6,2,0,0,1,1}{7,3,0} + 
\alpha_{9}\Ib{1}{0} \Ib{1}{0} \Ib{3}{0} + 
\alpha_{10}\Ib{1}{0} \Ib{2}{0} \Ib{2}{0} 
\Big]+ \nn
+\CA^2 \Nf \Big[ &%
\alpha_{11}\KK{1,1,0,0,2,1}{0,0,0} + 
\alpha_{12}\KK{1,1,0,0,2,2}{2,0,0} + 
\alpha_{13}\KK{1,1,0,0,3,1}{1,1,0} + 
\alpha_{14}\KK{1,1,0,0,3,1}{2,0,0} + \nn&%
\alpha_{15}\KK{2,1,0,0,1,1}{0,0,0} + 
\alpha_{16}\KK{2,1,0,0,2,1}{0,2,0} + 
\alpha_{17}\KK{2,1,0,0,2,1}{2,0,0} + 
\alpha_{18}\KK{2,1,0,0,3,1}{3,1,0} + \nn&%
\alpha_{19}\KK{2,2,0,0,1,1}{0,0,2} + 
\alpha_{20}\KK{2,2,0,0,1,1}{1,1,0} + 
\alpha_{21}\KK{2,2,0,0,1,1}{2,0,0} + 
\alpha_{22}\KK{3,1,0,0,1,1}{0,0,2} + \nn&%
\alpha_{23}\KK{3,1,0,0,1,1}{0,2,0} + 
\alpha_{24}\KK{3,1,0,0,1,1}{1,1,0} + 
\alpha_{25}\KK{3,1,0,0,1,1}{2,0,0} + 
\alpha_{26}\KK{3,1,0,0,2,1}{1,3,0} + \nn&%
\alpha_{27}\KK{3,2,0,0,1,1}{0,4,0} + 
\alpha_{28}\KK{4,1,0,0,1,1}{1,1,2} + 
\alpha_{29}\KK{4,1,0,0,1,1}{1,3,0} + 
\alpha_{30}\KK{4,1,0,0,1,1}{2,2,0} + \nn&%
\alpha_{31}\KK{4,1,0,0,1,1}{4,0,0} + 
\alpha_{32}\KK{4,2,0,0,1,1}{6,0,0} + 
\alpha_{33}\KK{5,1,0,0,1,1}{3,3,0} + 
\alpha_{34}\KK{5,1,0,0,1,1}{4,0,2} + \nn&%
\alpha_{35}\KK{5,1,0,0,1,1}{5,1,0} + 
\alpha_{36}\KK{5,1,0,0,1,1}{6,0,0} + 
\alpha_{37}\KK{6,1,0,0,1,1}{7,1,0} + 
\alpha_{38}\KK{6,1,0,0,1,1}{8,0,0} + \nn&%
\alpha_{39}\KK{6,1,0,0,2,1}{7,0,3} + 
\alpha_{40}\KK{6,1,0,0,2,1}{9,0,1} + 
\alpha_{41}\KK{6,1,0,0,2,1}{10,0,0} + 
\alpha_{42}\KK{6,2,0,0,1,1}{7,3,0} + \nn&%
\alpha_{43}\KK{6,2,0,0,1,1}{8,2,0} + 
\alpha_{44}\KK{7,1,0,0,1,1}{8,0,2} + 
\alpha_{45}\KK{7,1,0,0,1,1}{8,2,0} + 
\alpha_{46}\KK{7,1,0,0,1,1}{9,0,1} + \nn&%
\alpha_{47}\KK{7,1,0,0,1,1}{9,1,0} + 
\alpha_{48}\KK{7,1,0,0,1,1}{10,0,0} + 
\alpha_{49}\LL{2,1,0,0,1,1}{0,0,0} + 
\alpha_{50}\LL{3,1,0,0,1,1}{2,0,0} + \nn&%
\alpha_{51}\If{1}{0} \If{1}{0} \Ib{3}{0} + 
\alpha_{52}\If{1}{0} \Ib{2}{0} \If{2}{0} + 
\alpha_{53}\If{1}{0} \Ib{2}{0} \Ib{2}{0} + 
\alpha_{54}\If{1}{2} \If{2}{0} \Ib{3}{0} + 
\alpha_{55}\Ib{1}{0} \If{1}{0} \If{3}{0} + \nn&%
\alpha_{56}\Ib{1}{0} \If{1}{0} \Ib{3}{0} + 
\alpha_{57}\Ib{1}{0} \If{2}{0} \If{2}{0} + 
\alpha_{58}\Ib{1}{0} \Ib{1}{0} \If{3}{0} + 
\alpha_{59}\Ib{1}{0} \Ib{2}{0} \If{2}{0} + 
\alpha_{60}\Ib{1}{2} \If{2}{0} \Ib{3}{0} 
\Big]+ \nn
+\CA \Nf^2 \Big[ &%
\alpha_{61}\LL{2,1,0,0,1,1}{0,0,0} + 
\alpha_{62}\LL{2,2,0,0,1,1}{0,0,2} + 
\alpha_{63}\LL{3,1,0,0,1,1}{0,2,0} + 
\alpha_{64}\LL{3,1,0,0,1,1}{2,0,0} + \nn&%
\alpha_{65}\LL{4,1,0,0,1,1}{1,3,0} + 
\alpha_{66}\LL{5,1,0,0,1,1}{6,0,0} + 
\alpha_{67}\LL{5,3,0,0,1,1}{6,4,0} + 
\alpha_{68}\LL{6,2,0,0,1,1}{7,3,0} + \nn&%
\alpha_{69}\If{1}{0} \If{1}{0} \If{3}{0} + 
\alpha_{70}\If{1}{0} \If{1}{0} \Ib{3}{0} + 
\alpha_{71}\If{1}{0} \If{2}{0} \If{2}{0} + 
\alpha_{72}\If{1}{0} \Ib{2}{0} \If{2}{0} 
\Big]+ \nn
+\Nf^2 \CF \Big[ &%
\alpha_{73}\LL{2,1,0,0,1,1}{0,0,0} + 
\alpha_{74}\LL{2,2,0,0,1,1}{0,0,2} + 
\alpha_{75}\LL{3,1,0,0,1,1}{0,2,0} + 
\alpha_{76}\LL{3,1,0,0,1,1}{2,0,0} + \nn&%
\alpha_{77}\LL{5,1,0,0,1,1}{6,0,0} + 
\alpha_{78}\If{1}{0} \If{1}{0} \If{3}{0} + 
\alpha_{79}\If{1}{0} \If{2}{0} \If{2}{0} + 
\alpha_{80}\If{1}{0} \Ib{2}{0} \If{2}{0} 
\Big]+ \nn
+\Nf \CF^2 \Big[ &%
\alpha_{81}\KK{1,1,0,0,2,1}{0,0,0} + 
\alpha_{82}\KK{1,1,0,0,2,2}{2,0,0} + 
\alpha_{83}\KK{1,1,0,0,3,1}{1,1,0} + 
\alpha_{84}\KK{2,1,0,0,1,1}{0,0,0} + \nn&%
\alpha_{85}\KK{2,1,0,0,2,1}{0,2,0} + 
\alpha_{86}\KK{2,1,0,0,2,1}{2,0,0} + 
\alpha_{87}\KK{2,2,0,0,1,1}{1,1,0} + 
\alpha_{88}\KK{2,2,0,0,1,1}{2,0,0} + \nn&%
\alpha_{89}\KK{3,1,0,0,1,1}{0,2,0} + 
\alpha_{90}\KK{3,1,0,0,1,1}{1,1,0} + 
\alpha_{91}\KK{3,1,0,0,1,1}{2,0,0} + 
\alpha_{92}\KK{3,1,0,0,2,1}{1,3,0} + \nn&%
\alpha_{93}\KK{3,2,0,0,1,1}{0,4,0} + 
\alpha_{94}\KK{4,1,0,0,1,1}{1,3,0} + 
\alpha_{95}\KK{4,1,0,0,1,1}{2,2,0} + 
\alpha_{96}\KK{4,1,0,0,1,1}{4,0,0} + \nn&%
\alpha_{97}\KK{4,2,0,0,1,1}{6,0,0} + 
\alpha_{98}\KK{5,1,0,0,1,1}{3,3,0} + 
\alpha_{99}\KK{5,1,0,0,1,1}{5,1,0} + 
\alpha_{100}\LL{2,1,0,0,1,1}{0,0,0} + \nn&%
\alpha_{101}\LL{3,1,0,0,1,1}{2,0,0} + 
\alpha_{102}\If{1}{0} \If{1}{0} \If{3}{0} + 
\alpha_{103}\If{1}{0} \If{1}{0} \Ib{3}{0} + 
\alpha_{104}\If{1}{0} \If{2}{0} \If{2}{0} + \nn&%
\alpha_{105}\If{1}{0} \Ib{2}{0} \If{2}{0} + 
\alpha_{106}\If{1}{2} \If{2}{0} \Ib{3}{0} + 
\alpha_{107}\Ib{1}{0} \If{1}{0} \If{3}{0} + 
\alpha_{108}\Ib{1}{0} \If{2}{0} \If{2}{0} + 
\alpha_{109}\Ib{1}{0} \Ib{1}{0} \If{3}{0} + \nn&%
\alpha_{110}\Ib{1}{0} \Ib{2}{0} \If{2}{0} + 
\alpha_{111}\Ib{1}{2} \If{2}{0} \Ib{3}{0} 
\Big]+ \nn
+\CA \Nf \CF \Big[ &%
\alpha_{112}\KK{1,1,0,0,2,1}{0,0,0} + 
\alpha_{113}\KK{1,1,0,0,2,2}{2,0,0} + 
\alpha_{114}\KK{2,1,0,0,1,1}{0,0,0} + 
\alpha_{115}\KK{2,1,0,0,2,1}{0,2,0} + \nn&%
\alpha_{116}\KK{2,1,0,0,2,1}{2,0,0} + 
\alpha_{117}\KK{2,2,0,0,1,1}{1,1,0} + 
\alpha_{118}\KK{2,2,0,0,1,1}{2,0,0} + 
\alpha_{119}\KK{3,1,0,0,1,1}{0,2,0} + \nn&%
\alpha_{120}\KK{3,1,0,0,1,1}{1,1,0} + 
\alpha_{121}\KK{3,1,0,0,1,1}{2,0,0} + 
\alpha_{122}\KK{3,1,0,0,2,1}{1,3,0} + 
\alpha_{123}\KK{3,2,0,0,1,1}{0,4,0} + \nn&%
\alpha_{124}\KK{4,1,0,0,1,1}{1,3,0} + 
\alpha_{125}\KK{4,1,0,0,1,1}{2,2,0} + 
\alpha_{126}\KK{4,1,0,0,1,1}{4,0,0} + 
\alpha_{127}\KK{4,2,0,0,1,1}{6,0,0} + \nn&%
\alpha_{128}\KK{5,1,0,0,1,1}{3,3,0} + 
\alpha_{129}\LL{2,1,0,0,1,1}{0,0,0} + 
\alpha_{130}\LL{3,1,0,0,1,1}{2,0,0} + 
\alpha_{131}\If{1}{0} \If{1}{0} \Ib{3}{0} + \nn&%
\alpha_{132}\If{1}{0} \If{2}{0} \If{2}{0} + 
\alpha_{133}\If{1}{0} \Ib{2}{0} \If{2}{0} + 
\alpha_{134}\If{1}{2} \If{2}{0} \Ib{3}{0} + 
\alpha_{135}\Ib{1}{0} \If{2}{0} \If{2}{0} + 
\alpha_{136}\Ib{1}{0} \Ib{2}{0} \If{2}{0} + \nn&%
\alpha_{137}\Ib{1}{2} \If{2}{0} \Ib{3}{0} 
\Big] \;,
\end{align}
\begin{align}
\la{eq:betas}
\PiTcp = \CA^3 \Big[ &%
\beta_{1}\JJ{2,2,0,0,1,1}{0,0,0} + 
\beta_{2}\JJ{3,1,0,0,1,1}{0,0,0} + 
\beta_{3}\JJ{3,2,0,0,1,1}{0,0,2} + 
\beta_{4}\JJ{4,1,0,0,1,1}{0,2,0} + \nn&%
\beta_{5}\JJ{5,1,0,0,1,1}{2,2,0} + 
\beta_{6}\JJ{5,1,0,0,1,1}{4,0,0} + 
\beta_{7}\JJ{7,1,0,0,1,1}{8,0,0} + 
\beta_{8}\JJ{7,3,0,-1,1,1}{7,3,0} + \nn&%
\beta_{9}\JJ{8,2,0,-1,1,1}{8,2,0} + 
\beta_{10}\Ib{1}{0} \Ib{1}{0} \Ib{4}{0} + 
\beta_{11}\Ib{1}{0} \Ib{2}{0} \Ib{3}{0} + 
\beta_{12}\Ib{2}{0} \Ib{2}{0} \Ib{2}{0} 
\Big]+ \nn
+\CA^2 \Nf \Big[ &%
\beta_{13}\KK{1,1,0,0,2,2}{0,0,0} + 
\beta_{14}\KK{1,1,0,0,3,1}{0,0,0} + 
\beta_{15}\KK{1,1,0,0,3,2}{2,0,0} + 
\beta_{16}\KK{1,1,0,0,4,1}{1,1,0} + \nn&%
\beta_{17}\KK{1,1,0,0,4,1}{2,0,0} + 
\beta_{18}\KK{2,1,0,0,2,1}{0,0,0} + 
\beta_{19}\KK{2,1,0,0,3,1}{0,2,0} + 
\beta_{20}\KK{2,1,0,0,3,1}{2,0,0} + \nn&%
\beta_{21}\KK{2,1,0,0,4,1}{4,0,0} + 
\beta_{22}\KK{2,2,0,0,1,1}{0,0,0} + 
\beta_{23}\KK{3,1,0,0,1,1}{0,0,0} + 
\beta_{24}\KK{3,1,0,0,2,1}{0,2,0} + \nn&%
\beta_{25}\KK{3,1,0,0,2,1}{1,1,0} + 
\beta_{26}\KK{3,1,0,0,3,1}{1,3,0} + 
\beta_{27}\KK{3,1,0,0,3,1}{4,0,0} + 
\beta_{28}\KK{3,2,0,0,1,1}{0,0,2} + \nn&%
\beta_{29}\KK{3,2,0,0,1,1}{0,2,0} + 
\beta_{30}\KK{3,2,0,0,1,1}{1,1,0} + 
\beta_{31}\KK{4,1,0,0,1,1}{0,0,2} + 
\beta_{32}\KK{4,1,0,0,1,1}{0,2,0} + \nn&%
\beta_{33}\KK{4,1,0,0,1,1}{1,1,0} + 
\beta_{34}\KK{4,1,0,0,1,1}{2,0,0} + 
\beta_{35}\KK{4,1,0,0,2,1}{1,3,0} + 
\beta_{36}\KK{4,2,0,0,1,1}{0,4,0} + \nn&%
\beta_{37}\KK{4,2,0,0,1,1}{4,0,0} + 
\beta_{38}\KK{5,1,0,0,1,1}{1,1,2} + 
\beta_{39}\KK{5,1,0,0,1,1}{1,3,0} + 
\beta_{40}\KK{5,1,0,0,1,1}{2,0,2} + \nn&%
\beta_{41}\KK{5,1,0,0,1,1}{2,2,0} + 
\beta_{42}\KK{5,1,0,0,1,1}{3,1,0} + 
\beta_{43}\KK{5,1,0,0,1,1}{4,0,0} + 
\beta_{44}\KK{6,1,0,0,1,1}{3,3,0} + \nn&%
\beta_{45}\KK{6,1,0,0,1,1}{4,2,0} + 
\beta_{46}\KK{6,1,0,0,1,1}{6,0,0} + 
\beta_{47}\KK{6,2,0,0,1,1}{8,0,0} + 
\beta_{48}\KK{7,1,0,0,1,1}{5,3,0} + \nn&%
\beta_{49}\KK{7,1,0,0,1,1}{6,0,2} + 
\beta_{50}\KK{7,1,0,0,1,1}{7,1,0} + 
\beta_{51}\KK{7,1,0,0,1,1}{8,0,0} + 
\beta_{52}\KK{7,2,0,0,1,1}{8,2,0} + \nn&%
\beta_{53}\KK{7,2,0,0,1,1}{9,1,0} + 
\beta_{54}\KK{8,1,-1,0,2,1}{8,0,2} + 
\beta_{55}\KK{8,1,-1,0,2,1}{10,0,0} + 
\beta_{56}\KK{8,1,0,0,1,1}{8,0,2} + \nn&%
\beta_{57}\KK{8,1,0,0,1,1}{9,0,1} + 
\beta_{58}\KK{8,1,0,0,1,1}{9,1,0} + 
\beta_{59}\KK{8,1,0,0,1,1}{10,0,0} + 
\beta_{60}\KK{8,2,-1,0,1,1}{8,1,1} + \nn&%
\beta_{61}\KK{8,2,-1,0,1,1}{9,1,0} + 
\beta_{62}\LL{2,2,0,0,1,1}{0,0,0} + 
\beta_{63}\LL{3,1,0,0,1,1}{0,0,0} + 
\beta_{64}\LL{5,1,0,0,1,1}{4,0,0} + \nn&%
\beta_{65}\If{1}{0} \If{1}{0} \Ib{4}{0} + 
\beta_{66}\If{1}{0} \If{2}{0} \Ib{3}{0} + 
\beta_{67}\If{1}{0} \Ib{2}{0} \If{3}{0} + 
\beta_{68}\If{1}{0} \Ib{2}{0} \Ib{3}{0} + 
\beta_{69}\If{1}{2} \If{2}{0} \Ib{4}{0} + \nn&%
\beta_{70}\If{1}{2} \Ib{2}{0} \If{4}{0} + 
\beta_{71}\If{1}{2} \Ib{3}{0} \If{3}{0} + 
\beta_{72}\If{2}{0} \If{2}{0} \If{2}{0} + 
\beta_{73}\Ib{1}{0} \If{1}{0} \If{4}{0} + 
\beta_{74}\Ib{1}{0} \If{1}{0} \Ib{4}{0} + \nn&%
\beta_{75}\Ib{1}{0} \If{2}{0} \If{3}{0} + 
\beta_{76}\Ib{1}{0} \If{2}{0} \Ib{3}{0} + 
\beta_{77}\Ib{1}{0} \Ib{1}{0} \If{4}{0} + 
\beta_{78}\Ib{1}{0} \Ib{2}{0} \If{3}{0} + 
\beta_{79}\Ib{1}{2} \If{2}{0} \Ib{4}{0} + \nn&%
\beta_{80}\Ib{1}{2} \Ib{2}{0} \If{4}{0} + 
\beta_{81}\Ib{1}{2} \Ib{3}{0} \If{3}{0} + 
\beta_{82}\Ib{2}{0} \If{2}{0} \If{2}{0} + 
\beta_{83}\Ib{2}{0} \Ib{2}{0} \If{2}{0} 
\Big]+ \nn
+\CA \Nf^2 \Big[ &%
\beta_{84}\LL{2,2,0,0,1,1}{0,0,0} + 
\beta_{85}\LL{3,1,0,0,1,1}{0,0,0} + 
\beta_{86}\LL{3,2,0,0,1,1}{0,0,2} + 
\beta_{87}\LL{4,1,0,0,1,1}{0,2,0} + \nn&%
\beta_{88}\LL{5,1,0,0,1,1}{2,2,0} + 
\beta_{89}\LL{5,1,0,0,1,1}{4,0,0} + 
\beta_{90}\LL{7,1,0,0,1,1}{8,0,0} + 
\beta_{91}\LL{7,3,0,-1,1,1}{7,3,0} + \nn&%
\beta_{92}\LL{8,2,0,-1,1,1}{8,2,0} + 
\beta_{93}\If{1}{0} \If{1}{0} \If{4}{0} + 
\beta_{94}\If{1}{0} \If{1}{0} \Ib{4}{0} + 
\beta_{95}\If{1}{0} \If{2}{0} \If{3}{0} + 
\beta_{96}\If{1}{0} \If{2}{0} \Ib{3}{0} + \nn&%
\beta_{97}\If{1}{0} \Ib{2}{0} \If{3}{0} + 
\beta_{98}\If{2}{0} \If{2}{0} \If{2}{0} + 
\beta_{99}\Ib{2}{0} \If{2}{0} \If{2}{0} 
\Big]+ \nn
+\Nf^2 \CF \Big[ &%
\beta_{100}\LL{2,2,0,0,1,1}{0,0,0} + 
\beta_{101}\LL{3,1,0,0,1,1}{0,0,0} + 
\beta_{102}\LL{3,2,0,0,1,1}{0,0,2} + 
\beta_{103}\LL{4,1,0,0,1,1}{0,2,0} + \nn&%
\beta_{104}\LL{5,1,0,0,1,1}{2,2,0} + 
\beta_{105}\LL{5,1,0,0,1,1}{4,0,0} + 
\beta_{106}\LL{7,1,0,0,1,1}{8,0,0} + 
\beta_{107}\If{1}{0} \If{1}{0} \If{4}{0} + \nn&%
\beta_{108}\If{1}{0} \If{2}{0} \If{3}{0} +
\beta_{109}\If{1}{0} \If{2}{0} \Ib{3}{0} + 
\beta_{110}\If{1}{0} \Ib{2}{0} \If{3}{0} + 
\beta_{111}\If{2}{0} \If{2}{0} \If{2}{0} 
\Big]+ \nn
+\Nf \CF^2 \Big[ &%
\beta_{112}\KK{1,1,0,0,2,2}{0,0,0} + 
\beta_{113}\KK{1,1,0,0,3,1}{0,0,0} + 
\beta_{114}\KK{1,1,0,0,4,1}{1,1,0} + 
\beta_{115}\KK{1,1,0,0,4,1}{2,0,0} + \nn&%
\beta_{116}\KK{2,1,0,0,2,1}{0,0,0} + 
\beta_{117}\KK{2,1,0,0,3,1}{0,2,0} + 
\beta_{118}\KK{2,1,0,0,3,1}{2,0,0} + 
\beta_{119}\KK{2,2,0,0,1,1}{0,0,0} + \nn&%
\beta_{120}\KK{3,1,0,0,1,1}{0,0,0} + 
\beta_{121}\KK{3,1,0,0,2,1}{0,2,0} + 
\beta_{122}\KK{3,1,0,0,2,1}{1,1,0} + 
\beta_{123}\KK{3,1,0,0,3,1}{1,3,0} + \nn&%
\beta_{124}\KK{3,1,0,0,3,1}{4,0,0} + 
\beta_{125}\KK{3,2,0,0,1,1}{0,2,0} + 
\beta_{126}\KK{3,2,0,0,1,1}{1,1,0} + 
\beta_{127}\KK{4,1,0,0,1,1}{0,2,0} + \nn&%
\beta_{128}\KK{4,1,0,0,1,1}{1,1,0} + 
\beta_{129}\KK{4,1,0,0,1,1}{2,0,0} + 
\beta_{130}\KK{4,1,0,0,2,1}{1,3,0} + 
\beta_{131}\KK{4,2,0,0,1,1}{0,4,0} + \nn&%
\beta_{132}\KK{4,2,0,0,1,1}{4,0,0} + 
\beta_{133}\KK{5,1,0,0,1,1}{1,3,0} + 
\beta_{134}\KK{5,1,0,0,1,1}{2,2,0} + 
\beta_{135}\KK{5,1,0,0,1,1}{3,1,0} + \nn&%
\beta_{136}\KK{5,1,0,0,1,1}{4,0,0} + 
\beta_{137}\KK{6,1,0,0,1,1}{3,3,0} + 
\beta_{138}\KK{6,1,0,0,1,1}{4,2,0} + 
\beta_{139}\KK{6,1,0,0,1,1}{6,0,0} + \displaybreak\nn&%
\beta_{140}\KK{6,2,0,0,1,1}{8,0,0} + 
\beta_{141}\KK{7,1,0,0,1,1}{5,3,0} + 
\beta_{142}\KK{7,1,0,0,1,1}{7,1,0} + 
\beta_{143}\LL{2,2,0,0,1,1}{0,0,0} + \nn&%
\beta_{144}\LL{3,1,0,0,1,1}{0,0,0} + 
\beta_{145}\LL{5,1,0,0,1,1}{4,0,0} + 
\beta_{146}\If{1}{0} \If{1}{0} \If{4}{0} + 
\beta_{147}\If{1}{0} \If{1}{0} \Ib{4}{0} + \nn&%
\beta_{148}\If{1}{0} \If{2}{0} \If{3}{0} + 
\beta_{149}\If{1}{0} \If{2}{0} \Ib{3}{0} + 
\beta_{150}\If{1}{0} \Ib{2}{0} \If{3}{0} + 
\beta_{151}\If{1}{2} \If{2}{0} \Ib{4}{0} + 
\beta_{152}\If{1}{2} \Ib{3}{0} \If{3}{0} + \nn&%
\beta_{153}\If{2}{0} \If{2}{0} \If{2}{0} + 
\beta_{154}\Ib{1}{0} \If{1}{0} \If{4}{0} + 
\beta_{155}\Ib{1}{0} \If{2}{0} \If{3}{0} + 
\beta_{156}\Ib{1}{0} \If{2}{0} \Ib{3}{0} + 
\beta_{157}\Ib{1}{0} \Ib{1}{0} \If{4}{0} + \nn&%
\beta_{158}\Ib{1}{0} \Ib{2}{0} \If{3}{0} + 
\beta_{159}\Ib{1}{2} \If{2}{0} \Ib{4}{0} + 
\beta_{160}\Ib{1}{2} \Ib{3}{0} \If{3}{0} + 
\beta_{161}\Ib{2}{0} \If{2}{0} \If{2}{0} + 
\beta_{162}\Ib{2}{0} \Ib{2}{0} \If{2}{0} 
\Big]+ \nn
+\CA \Nf \CF \Big[ &%
\beta_{163}\KK{1,1,0,0,2,2}{0,0,0} + 
\beta_{164}\KK{1,1,0,0,3,1}{0,0,0} + 
\beta_{165}\KK{1,1,0,0,3,2}{2,0,0} + 
\beta_{166}\KK{1,1,0,0,4,1}{2,0,0} + \nn&%
\beta_{167}\KK{2,1,0,0,2,1}{0,0,0} + 
\beta_{168}\KK{2,1,0,0,3,1}{0,2,0} + 
\beta_{169}\KK{2,1,0,0,3,1}{2,0,0} + 
\beta_{170}\KK{2,2,0,0,1,1}{0,0,0} + \nn&%
\beta_{171}\KK{3,1,0,0,1,1}{0,0,0} + 
\beta_{172}\KK{3,1,0,0,2,1}{0,2,0} + 
\beta_{173}\KK{3,1,0,0,2,1}{1,1,0} + 
\beta_{174}\KK{3,1,0,0,3,1}{1,3,0} + \nn&%
\beta_{175}\KK{3,1,0,0,3,1}{4,0,0} + 
\beta_{176}\KK{3,2,0,0,1,1}{0,2,0} + 
\beta_{177}\KK{3,2,0,0,1,1}{1,1,0} + 
\beta_{178}\KK{4,1,0,0,1,1}{0,2,0} + \nn&%
\beta_{179}\KK{4,1,0,0,1,1}{1,1,0} + 
\beta_{180}\KK{4,1,0,0,1,1}{2,0,0} + 
\beta_{181}\KK{4,1,0,0,2,1}{1,3,0} + 
\beta_{182}\KK{4,2,0,0,1,1}{0,4,0} + \nn&%
\beta_{183}\KK{4,2,0,0,1,1}{4,0,0} + 
\beta_{184}\KK{5,1,0,0,1,1}{1,3,0} + 
\beta_{185}\KK{5,1,0,0,1,1}{2,2,0} + 
\beta_{186}\KK{5,1,0,0,1,1}{3,1,0} + \nn&%
\beta_{187}\KK{5,1,0,0,1,1}{4,0,0} + 
\beta_{188}\KK{6,1,0,0,1,1}{3,3,0} + 
\beta_{189}\KK{6,1,0,0,1,1}{4,2,0} + 
\beta_{190}\KK{6,1,0,0,1,1}{6,0,0} + \nn&%
\beta_{191}\KK{6,2,0,0,1,1}{8,0,0} + 
\beta_{192}\KK{7,1,0,0,1,1}{5,3,0} + 
\beta_{193}\LL{2,2,0,0,1,1}{0,0,0} + 
\beta_{194}\LL{3,1,0,0,1,1}{0,0,0} + \nn&%
\beta_{195}\LL{5,1,0,0,1,1}{4,0,0} + 
\beta_{196}\If{1}{0} \If{1}{0} \Ib{4}{0} + 
\beta_{197}\If{1}{0} \If{2}{0} \If{3}{0} + 
\beta_{198}\If{1}{0} \If{2}{0} \Ib{3}{0} + 
\beta_{199}\If{1}{0} \Ib{2}{0} \If{3}{0} + \nn&%
\beta_{200}\If{1}{2} \If{2}{0} \Ib{4}{0} + 
\beta_{201}\If{1}{2} \Ib{3}{0} \If{3}{0} + 
\beta_{202}\If{2}{0} \If{2}{0} \If{2}{0} + 
\beta_{203}\Ib{1}{0} \If{1}{0} \If{4}{0} + 
\beta_{204}\Ib{1}{0} \If{2}{0} \If{3}{0} + \nn&%
\beta_{205}\Ib{1}{0} \If{2}{0} \Ib{3}{0} + 
\beta_{206}\Ib{1}{0} \Ib{2}{0} \If{3}{0} + 
\beta_{207}\Ib{1}{2} \If{2}{0} \Ib{4}{0} + 
\beta_{208}\Ib{1}{2} \Ib{3}{0} \If{3}{0} + 
\beta_{209}\Ib{2}{0} \If{2}{0} \If{2}{0} + \nn&%
\beta_{210}\Ib{2}{0} \Ib{2}{0} \If{2}{0} 
\Big] \;.
\end{align}

Looking at the master integrals that are needed for the above two 
lengthy expressions let us note that, while most of them have
factors of $P_0$ etc.\ in the numerator,
only eight of them (those multiplying $\beta_{\{8,9,54,55,60,61,91,92\}}$) 
contain irreducible scalar products in the numerator and hence need
methods for their evaluation that go beyond those presented in 
Appendix \ref{se:mintegrals} 
(see, however, Ref.~\cite{Arnold:1994ps,Arnold:1994eb}, 
where examples of such sum-integrals were treated).
Also, some of the masters (such as e.g.\ those multiplying 
$\alpha_{\{7,8\}},\beta_{\{8,9\}}$) involve somewhat large
powers of propagators, which is a consequence of our ordering
prescription.
However, as was shown in App.~\ref{se:mintegrals} in terms of 
the generic power $N$, this does not seem to be a particularly
difficult obstacle.

We refrain from listing the coefficients $\alpha_{1...137}$
and $\beta_{1...210}$ here. 
They have the general form $\sum_{n}\xi^n\,p_n(d)/q_n(d)$, 
where $\xi$ is the gauge parameter (see \eq\nr{eq:propag})
and $p,q$ are polynomials in $d$.
The full expressions for \eqs\nr{eq:alphas} and \nr{eq:betas} are provided 
in computer-readable form on \cite{webpage:3loopTaylorCoeffs}.

%

\end{document}